\documentclass[a4paper,fleqn,usenatbib]{mnras}

\usepackage[T1]{fontenc}
\usepackage{ae,aecompl}

\usepackage{graphicx}	
\usepackage{amsmath}	

\usepackage{amssymb}	

\title[Carbon X-ray ISM absorption]{Carbon X-ray absorption in the local ISM: fingerprints in X-ray Novae spectra }

\author[Gatuzz et al.]{
Efra\'in~Gatuzz$^{1}$\thanks{E-mail: egatuzzs@eso.org},
J.-U. Ness$^{2}$,
T. W. Gorczyca$^{3}$,
M. F. Hasoglu$^{4}$,\newauthor
Timothy~R.~Kallman$^{5}$,
and  Javier~A.~Garc\'ia$^{6}$ 
\\
$^{1}$ESO, Karl-Schwarzschild-Strasse 2, D-85748 Garching bei M\"unchen, Germany\\
$^{2}$XMM-Newton Observatory SOC, European Space Astronomy Centre, Camino Bajo del Castillo s/n, Urb. Villafranca del Castillo,\\
 E-28692 Villanueva de la Ca\~nada, Madrid, Spain\\
$^{3}$Western Michigan University, Kalamazoo, MI 49008, USA\\
$^{4}$Department of Computer Engineering, Hasan Kalyoncu University, 27100 Sahinbey, Gaziantep, Turkey\\
$^{5}$NASA Goddard Space Flight Center, Greenbelt, MD 20771, USA\\
$^{6}$Cahill Center for Astronomy and Astrophysics, California Institute of Technology, Pasadena, CA 91125, USA\\
}

\date{Accepted XXX. Received YYY; in original form ZZZ}

\pubyear{2017}

\begin{document}
 \label{firstpage}
\pagerange{\pageref{firstpage}--\pageref{lastpage}}
\maketitle

\begin{abstract}
We present a study of the C K-edge using high-resolution LETGS {\it Chandra} spectra of four novae during their super-soft-source (SSS) phase. We identified absorption lines due to {\rm C}~{\sc ii} K$\alpha$, {\rm C}~{\sc iii} K$\alpha$ and {\rm C}~{\sc iii} K$\beta$ resonances. We used these astronomical observations to perform a benchmarking of the atomic data, which involves wavelength shifts of the resonances and photoionization cross-sections. We used improved atomic data to estimate the {\rm C}~{\sc ii} and {\rm C}~{\sc iii} column densities. The absence of physical shifts for the absorption lines, the consistence of the column densities between multiple observations and the high temperature required for the SSS nova atmosphere modeling support our conclusion about an ISM origin of the respective absorption lines.  Assuming a collisional ionization equilibrium plasma the maximum temperature derived from the ratio of {\rm C}~{\sc ii}/{\rm C}~{\sc iii} column densities of the absorbers correspond to $T_{max}< 3.05\times10^{4}$ K.
\end{abstract}

\begin{keywords}
ISM: structure -- ISM: atoms -- X-rays: ISM  
\end{keywords}

\section{Introduction}
High-resolution X-ray spectroscopy constitutes a powerful technique to study the elements associated with the local interstellar medium (ISM), defined as gas and dust between the stars. By using an X-ray bright source, acting as a lamp, the absorption features identified in the X-ray spectra provide information about the physical properties of the gas between the source and the observer. Using X-ray spectra of low mass X-ray binaries (LMXBs) the O, Fe, Ne, Mg and Si K absorption edges associated to the ISM have been analyzed in previous works \citep{jue04,ued05,jue06,yao09,pin10,pin13,cos12,gat13a,gat13b,lia13,luo14,gat14,gat15,sch16,gat16,nic16a,nic16b,joa16,gat18a}.

The ISM is composed of multiple phases which depends on their characteristic temperatures and densities. Carbon, which constitutes the fourth most abundant element in the Galaxy, can be used to probe the link between the different phases.  {\rm C}~{\sc i}, for example, has been used to analyze the cold Galactic gas which is characterized by a relatively low thermal pressure using the Space Telescope Imaging Spectrograph (STIS) on board the {\it Hubble Space Telescope}  \citep{jen01,bur10,jen11}, while  the {\rm C}~{\sc ii} 158 $\mu$m line allows the characterization of the cold atomic clouds in transition from atomic to molecular form \citep{pine13,lan14,pin14,pin17,ric17,sav17}. Also, it has been shown that solids that contain carbon atoms, such as graphite and polycyclic aromatic hydrocarbons, may constitute the main heat source for the ISM \citep{dra01,hel01,oka13,che17,sha18}. In this sense, it is essential to estimate the amount of {\rm C} depleted in the dust phase in order to fully understand the heating-cooling ISM processes.

One of the advantages of the high-resolution X-ray spectroscopy is that it provides access to both gas and solid components of the ISM. The C K-edge, located at 38-44~\AA\ wavelength, can be accessed only through the low-energy transmission grating (LETG) on board of the {\it Chandra} observatory.  Super-soft-sources (SSS) provide high count continuum spectra that can be used to perform such analysis. Although their X-ray spectra can be complex, showing spectral features due to multiple temperature components,  carbon absorption features have been identified \citep{nes09,rau10,van12,rau16}.

In this paper we present an analysis of the C absorption features in the ISM using {\it Chandra} high-resolution spectra of four sources. The outline of this paper is as follows: In Section~\ref{sec_dat} we describe the observations and data reduction. In Section~\ref{sec_ofit} we describe the C K-edge modeling and the atomic data involved. In Section~\ref{sec_dis} we discuss the main results. Finally, in Section~\ref{sec_con} we summarize our main conclusions. 
 \begin{table*}
\small
\caption{\label{tab1}List of {\it Chandra} LETGS-HRC observations.}
\centering
\begin{tabular}{clllclcc}
Source & ObsID.   & Obs. date & Exp. time& $N({\rm HI})$ & count-rate & Counts\\ 
& &&(ks)&($10^{21}$~cm$^{-2}$)&(counts/s)&(38--44~\AA)\\
\hline
\hline
\\
 KTEri&12097	  & 23-01-2010 &14.9 &0.52&11.52&25028\\
&12100	  &31-01-2010& 27.9&&77.73&18737\\
&12101	  &06-02-2010& 47.8&&37.48&16971\\
&12203	  &21-04-2010& 32.4&&106.5&26717 \\
Sgr2015b&16690	 &16-10-2015&48&1.11&12.07&106572 \\
&16691	 &12-11-2015&50&& 6.41&56768\\
V339Del&15742	 &09-11-2013&46&1.23&68.31&352919\\
&15743	 &06-12-2013&49&& 48.34&293788\\
V4743Sgr&3775	 &19-03-2003&20.3& 1.05&38.91&45173\\
&3776	 &18-07-2003&11.7&&37.16&26349\\
&4435	 &25-09-2003& 12.0&&19.78&16209\\ 
\hline
\multicolumn{7}{c}{$N({\rm HI})$ column densities obtained from \citet{kal05}.    }
\end{tabular}
\end{table*}
 \begin{table*}
\small
\caption{\label{tab2}Absorption line assignments with observed wavelength.}
\centering
\begin{tabular}{ccccccccccccccc}
Source & ObsID. & \multicolumn{3}{c}{${\rm CII}$} &  \multicolumn{2}{c}{${\rm CIII}$} \\
  &  & $K\alpha_{1}$ (\AA) & $K\alpha_{2}$ (\AA) & $K\alpha_{3}$ (\AA) & $K\alpha$ (\AA) & $K\beta$ (\AA)\\
\hline
\hline
\\
Theoretical& 	&$43.0592	 	$&$42.9867	 	$&$42.6818	 	$&$42.2360	 	$&$38.4459 $	\\
 KTEri&12097	&$	43.0599	\pm	0.0043	$&$	42.9653	\pm	0.0043	$&$	42.7838	\pm	0.0043	$&$	42.2091	\pm	0.0042	$&$	38.4100	\pm	0.0038$	\\
&12100	&$	43.0736	\pm	0.0039	$&$	43.0005	\pm	0.0039	$&$	42.7863	\pm	0.0039	$&--&--\\
&12101	&$	43.0623	\pm	0.0047	$&$	42.9875	\pm	0.0047	$&$	42.7626	\pm	0.0047	$&$	42.2042	\pm	0.0046	$&--\\
&12203	&$	43.0725	\pm	0.0047	$&$	43.0002	\pm	0.0047	$&$	42.7748	\pm	0.0047	$&--&--\\
Sgr2015b&16690	&$	43.0622	\pm	0.0034	$&$	42.9756	\pm	0.0034	$&$	42.7766	\pm	0.0034	$&--&--\\
&16691	&$	43.0887	\pm	0.0388	$&$	42.9777	\pm	0.0387	$&$	42.7830	\pm	0.0385	$&--&--\\
V339Del&15742	&$	43.0703	\pm	0.0043	$&$	42.9875	\pm	0.0043	$&$	42.7707	\pm	0.0043	$&$	42.1873	\pm	0.0042	$&$	38.4093	\pm	0.0038	$\\
&15743	&$	43.0642	\pm	0.0047	$&$	42.9906	\pm	0.0047	$&$	42.7800	\pm	0.0047	$&$	42.1752	\pm	0.0046	$&$	38.4082	\pm	0.0042	$\\
V4743Sgr&3775	&$	43.0725	\pm	0.0039	$&$	42.9847	\pm	0.0039	$&$	42.7666	\pm	0.0038	$&$	42.1878	\pm	0.0038	$&$	38.4039	\pm	0.0035	$\\
&3776	&$	43.0684	\pm	0.0043	$&$	42.9869	\pm	0.0043	$&$	42.7674	\pm	0.0043	$&$	42.1632	\pm	0.0042	$&--\\
&4435	&$	43.0684	\pm	0.0043	$&$	42.9795	\pm	0.0043	$&$	42.7748	\pm	0.0043	$&$	42.1615	\pm	0.0042	$&$	38.4133	\pm	0.0038	$\\
Average	&&$	43.0694	\pm	0.0074	$&$	42.9851	\pm	0.0074	$&$	42.7751	\pm	0.0074	$&$	42.1840\pm 0.0043	$&$	38.4089\pm 0.0038	$\\
\hline
\end{tabular}
\end{table*}

      \begin{figure}
\includegraphics[scale=0.42]{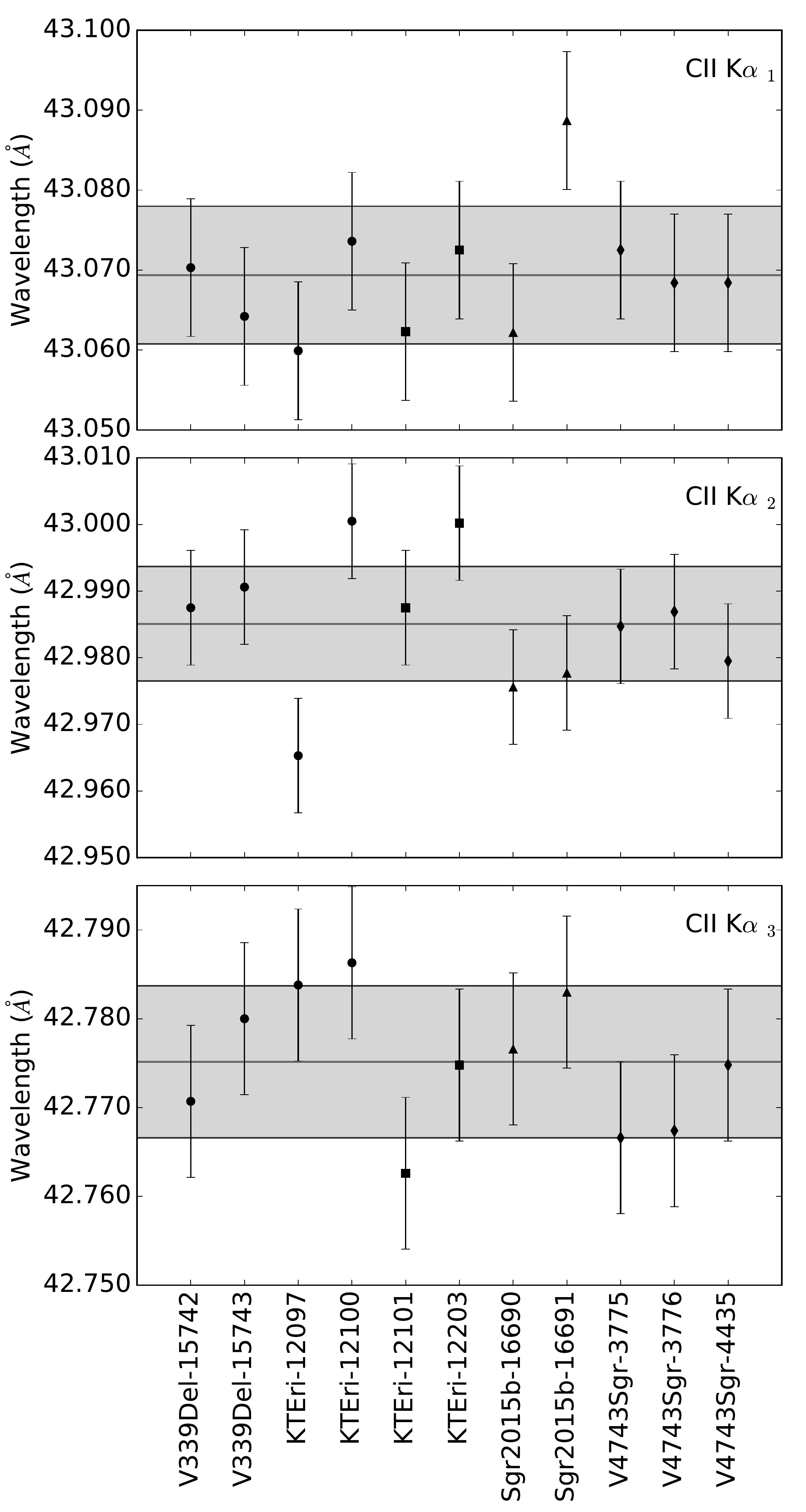}
      \caption{ {\rm C}~{\sc ii} K$\alpha$ resonances observational wavelengths determined from Gaussian fits. The horizontal grey region correspond to the average value and its 90$\%$ uncertainty.   }\label{fig1}
   \end{figure} 
    \begin{figure*}
\includegraphics[scale=0.46]{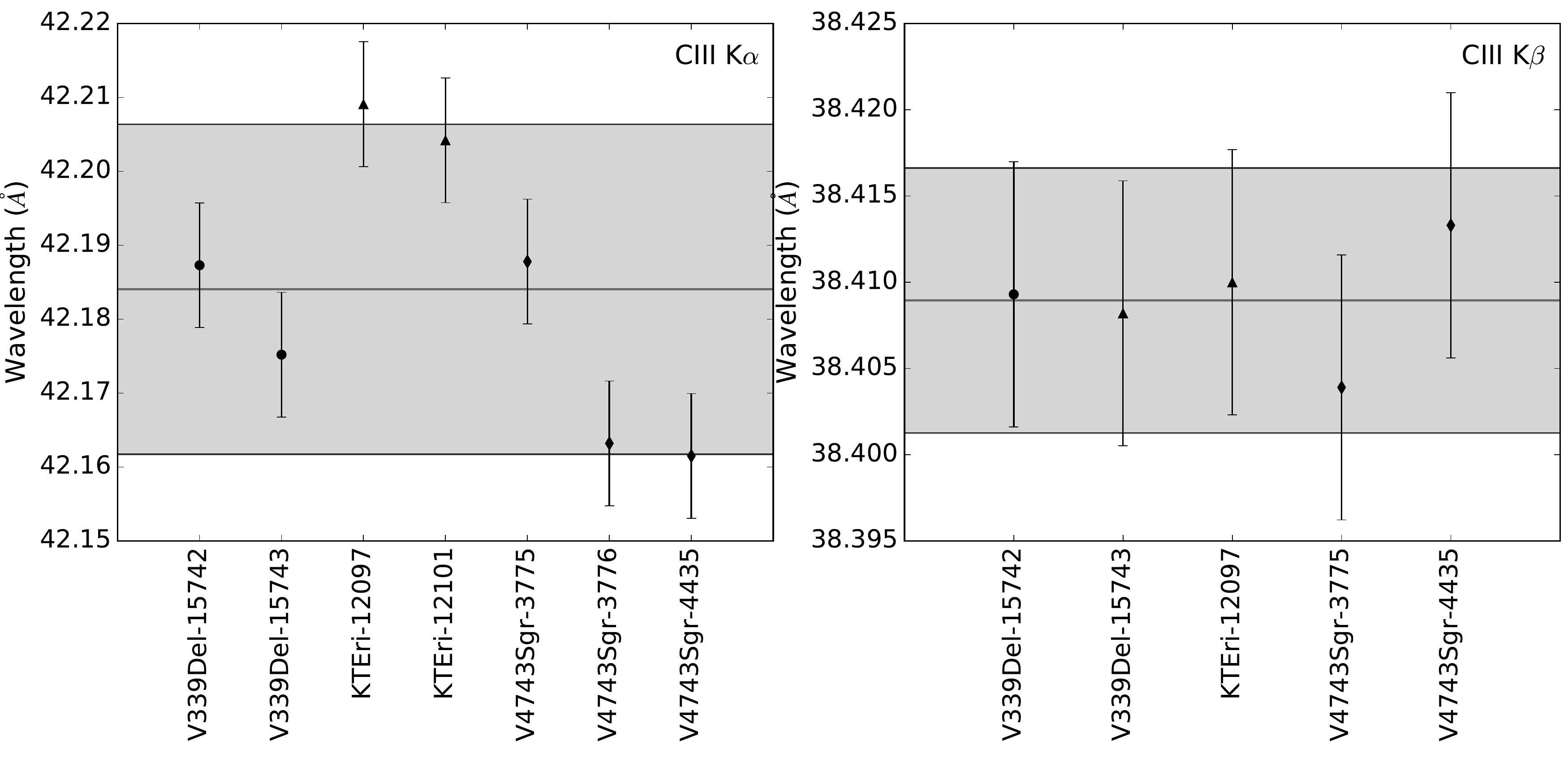}
      \caption{  Left pannel: {\rm C}~{\sc iii} K$\alpha$ resonance observational wavelengths determined from Gaussian fits. Right pannel: {\rm C}~{\sc iii} K$\beta$ resonance observational wavelengths determined from Gaussian fits. The horizontal gray region corresponds to the average value and its 90$\%$ uncertainty. }\label{fig2}
   \end{figure*} 
   
     \begin{figure*}
\includegraphics[scale=0.34]{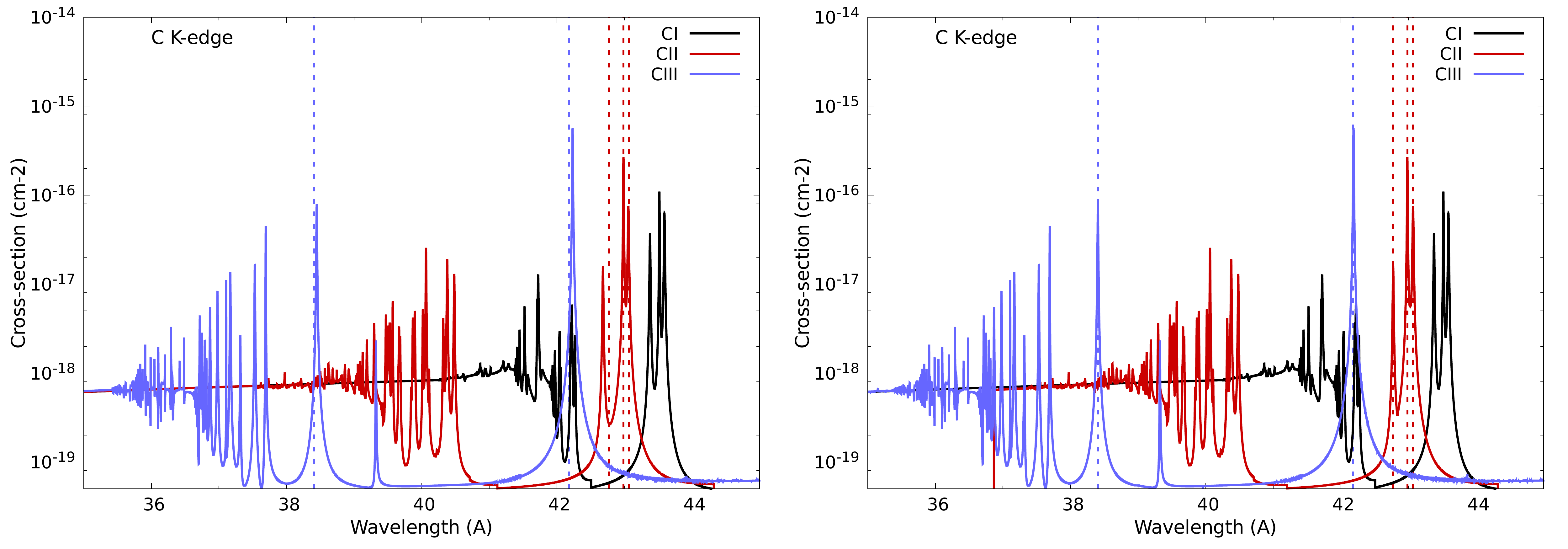}
      \caption{ {\rm C}~{\sc i}, {\rm C}~{\sc ii} and {\rm C}~{\sc iii} photoabsorption cross sections computed by Hasoglu et al. (2010) that are implemented in the {\tt ISMabs} model. Vertical dashed lines correspond to the average measurements listed in Table~\ref{tab1}. Left panel displays the original cross sections while right panel shows the same curves after the benchmarking. }\label{fig3}
   \end{figure*} 
\begin{figure*}
\includegraphics[scale=0.3]{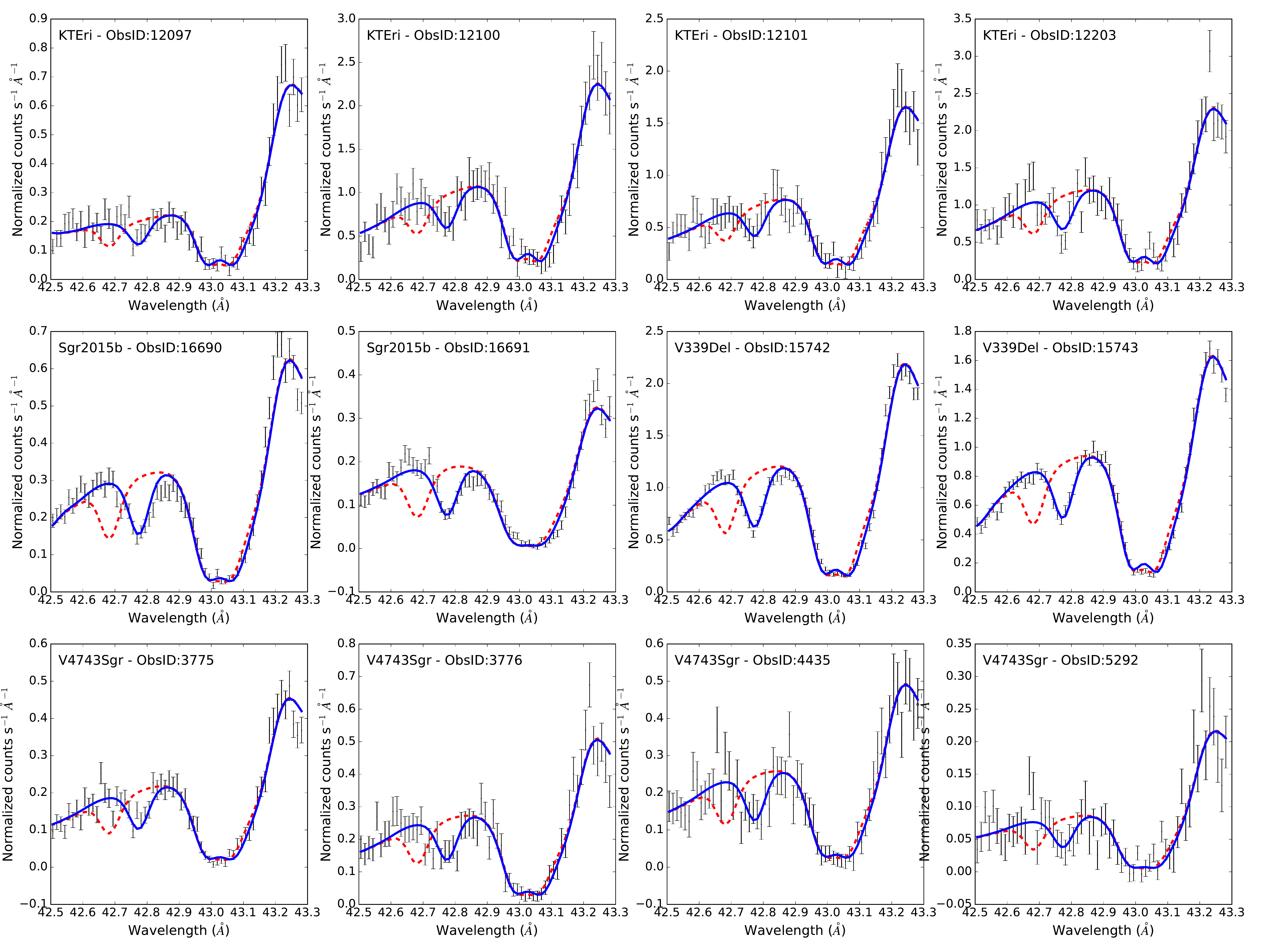}
      \caption{ Best fit results using {\it Chandra} LETG data for the {\rm C}~{\sc ii} K-edge wavelength region. Lines correspond to the model before and after the atomic data benchmarking (red dashed and blue solid lines, respectively) }\label{fig4}
   \end{figure*} 
   
\begin{table}
\small
\caption{\label{tab3}Column density best-fit results.}
\centering
\begin{tabular}{ccccccccccccccc}
Source & ObsID. & $N({\rm CII})$ &  $N({\rm CIII})$ & ${\rm CII}/{\rm CIII}$   \\
\hline
\hline
\\
 KTEri&12097	&$18.63	^{+	2.99	}_{-	2.55	} 	$&$	0.58^{+	0.58}_{-0.38} $  & $32.12\pm 27.00$ \\
&12100	&$19.51	^{	+3.34	}_{-	2.81	} 	$&$	 	$ \\
&12101	&$20.81	^{+	3.73	}_{-	3.10	}	 	$&$	 	$ \\
&12203	&$20.36	^{+	3.11	}_{	-2.65} 	$&$	 	$ \\
Sgr2015b&16690	&$33.50 ^{+	2.43	}_{-	2.19	}	 	$&$	 	$ \\
&16691	&$35.06^{	+4.27	}_{-	3.66	}	 	$&$	 	$ \\
V339Del&15742	&$24.98	^{+	0.87}_{-	0.83}	 	$&$0.53	^{+	0.14	}_{-	0.13	} $ & $47.13\pm 12.11$  \\
&15743	&$22.91	^{	+0.92	}_{-	0.88	}	 	$&$0.65	^{+	0.16}_{-	0.15	}	 	$ &$35.24\pm 8.51$ \\
V4743Sgr&3775	&$32.28	^{+	3.36	}_{-	2.93}	 	$&$	 	$ \\
&3776	&$29.50	^{+	4.64}_{-	3.81}	 	$&$	1.26	^{+	0.58	}_{-	0.46	} 	$   &$23.41\pm 10.24$ \\
&4435	&$31.19	^{+	6.40}_{	-4.98}	 	$&$	0.89	^{+	0.95	}_{-	0.60	} 	$ & $35.04\pm 31.37$  \\
\hline
\multicolumn{4}{c}{Column densities in units of $10^{16}$~cm$^{-2}$. }
\end{tabular}
\end{table}

        \begin{figure*}
\includegraphics[scale=0.38]{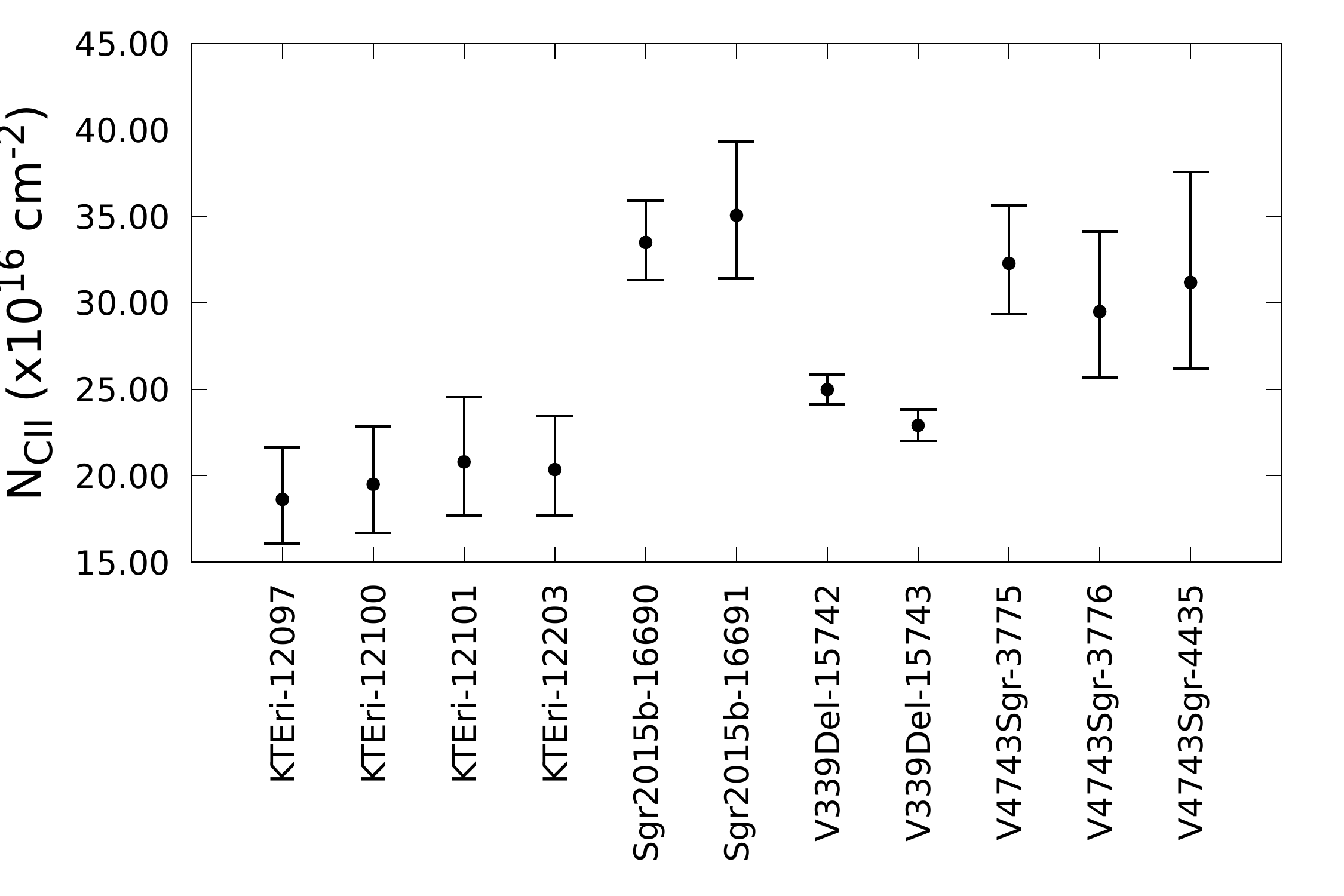} 
\includegraphics[scale=0.38]{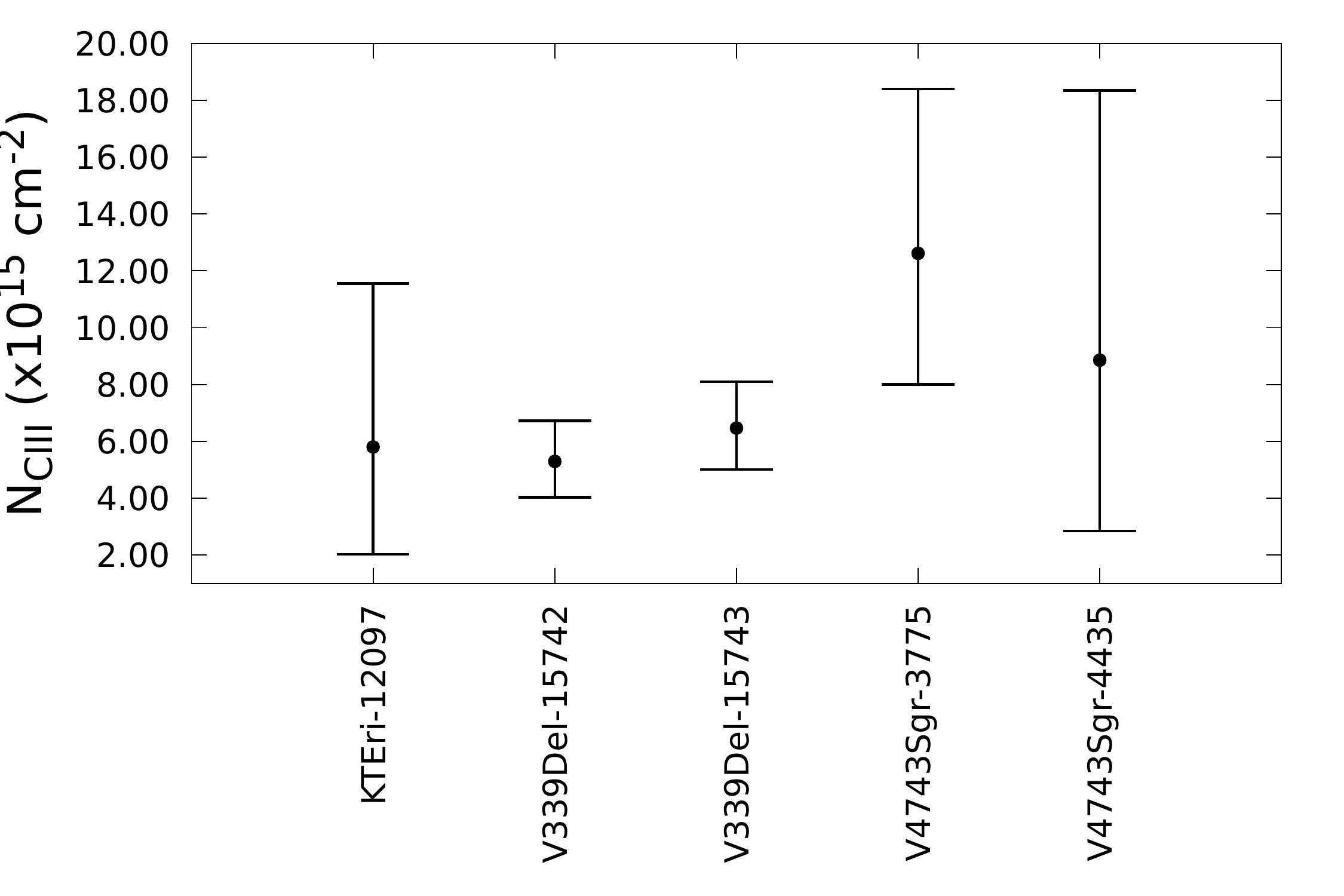} 
      \caption{   Left pannel: {\rm C}~{\sc ii} column densities obtained from the best {\tt ISMabs} fit.  Right pannel: {\rm C}~{\sc iii} column densities obtained from the best {\tt ISMabs} fit.  }\label{fig5}
   \end{figure*} 

\section{Observations and data reduction}\label{sec_dat}

We analyze {\it Chandra} spectra of four SSS in order to study the ISM carbon K-edge along different lines of sight. Because of their brightness and proximity, SSS high-resolution spectra constitute a useful way to analyze not only the binary system involved but also the ISM fingerprints identified as absorption features.   Table~\ref{tab1} lists the specifications for each LETGS-HRC observation,  the observation date, the exposure time, the hydrogen column density 21~cm measurement from the \citet{kal05} survey, the count rate and the number of counts in the C K-$\alpha$ wavelength region (38--44~\AA). The data were reduced using the Chandra Interactive Analysis of Observations software (CIAO,version 4.9) and following the standard procedure to obtain the Low-Energy Transmission Grating (LETG) spectra\footnote{\url{http://cxc.harvard.edu/ciao/threads/gspec.html}}. For each observation we combine the +1/-1 orders using the {\tt combine\_grating\_spectra} script. We use the {\sc xspec} analysis data package \citep[version 12.9.1\footnote{\url{https://heasarc.gsfc.nasa.gov/xanadu/xspec/}}]{arn96} to perform the spectral fitting. Finally, we use $\chi^{2}$ statistic with the weighting method for low counts regime defined by \citet{chu96}

\section{C K-edge modeling}\label{sec_ofit}
  
In order to analyze the C K-edge absorption region (38--44 \AA) we first used a functional model consisting of a power law continuum with absorption lines described by Gaussian profiles. Table~\ref{tab2} shows the wavelength position for all absorption lines identified in the spectra as well as their average values. Theoretical values, obtained from \citet{has10} calculations, are also listed. Both cameras, the High Resolution Camera (HRC) and the Advanced CCD Imaging Spectrometer (ACIS) have {\rm C}~{\sc i} instrumental absorption at $\sim43.6$~\AA\ due to absorption edges in the materials comprising the instruments, as is indicated by the Chandra Proposers' Observatory Guide\footnote{\url{http://cxc.harvard.edu/proposer/POG/html/index.html}}. We have not identified {\rm C}~{\sc i} absorption lines in excess of instrumental features that could be associated with the ISM in the X-ray spectra sample. In this sense, it has been shown that the {\rm C}~{\sc i} column density associated to the ISM tends to be lower than $10^{14}$ cm$^{-2}$ along multiple line-of-sights, including regions with large ${\rm HI}$ column densities \citep{jen11,ger15,wel16,pin17}.

Figure~\ref{fig1} shows the wavelength positions for each resonance  in the {\rm C}~{\sc ii}~K$\alpha$ triplet, which have been measured in all observations. It is important to note that, considering the uncertainties, the wavelength positions tend to agree not only between different observations of the same source but also for different sources. {\rm C}~{\sc iii}~K$\alpha$ and {\rm C}~{\sc iii}~K$\beta$ absorption lines were identified in 7 and 5 observations, respectively. Figure~\ref{fig2} shows the best-fit wavelength positions measured for both resonances. In both figures the horizontal gray regions indicate the average values and their uncertainties.  It is clear from the plots that, considering the uncertainties, the wavelength positions do not show significant shift between observations.     
      
We performed a benchmarking of the atomic data by comparing the observed and theoretical absorption lines in the C K-edge region. Left panel in Figure~\ref{fig3} shows the {\rm C}~{\sc i} (black line), {\rm C}~{\sc ii} (red line) and {\rm C}~{\sc iii} (blue line) photo-absorption cross sections computed by \citet{has10}. Vertical lines correspond to the average measurements listed in Table~\ref{tab1}. It is clear that the resonance positions for the K$\alpha$ transitions differ between the theoretical predictions and the observational measurements. In this sense we have adjusted cross sections in order to obtain the best possible agreement with the observed lines. The shifts are $+10.2$ m\AA, $-1.59$ m\AA, and $+93.3$ m\AA\ for the {\rm C}~{\sc ii} K$\alpha_{1}$, K$\alpha_{2}$ and K$\alpha_{3}$ resonances, respectively. The larger shift for the K$\alpha_{3}$ is expected because the $n=3$ resonances carry the greatest energy uncertainty along a Rydberg series \citep{has10}. For {\rm C}~{\sc iii} we move the K$\alpha$ and the whole cross-section by $-52$ m\AA, and $-37$ m\AA, respectively. From the theoretical point of view, a $\sim 59$ m\AA\ under-prediction in wavelength is expected \citep{has10}. Right panel in Figure~\ref{fig3} shows the cross sections after the wavelength corrections. It is important to mention that such benchmarking has been performed previously for the oxygen and neon photoabsorption cross-sections by \citet{gat13a,gat13b,gat15} concluding that the {\it Chandra} wavelength calibration can be safely used to correct the theoretical wavelength resonance positions.

           \begin{figure*}
\includegraphics[scale=0.3]{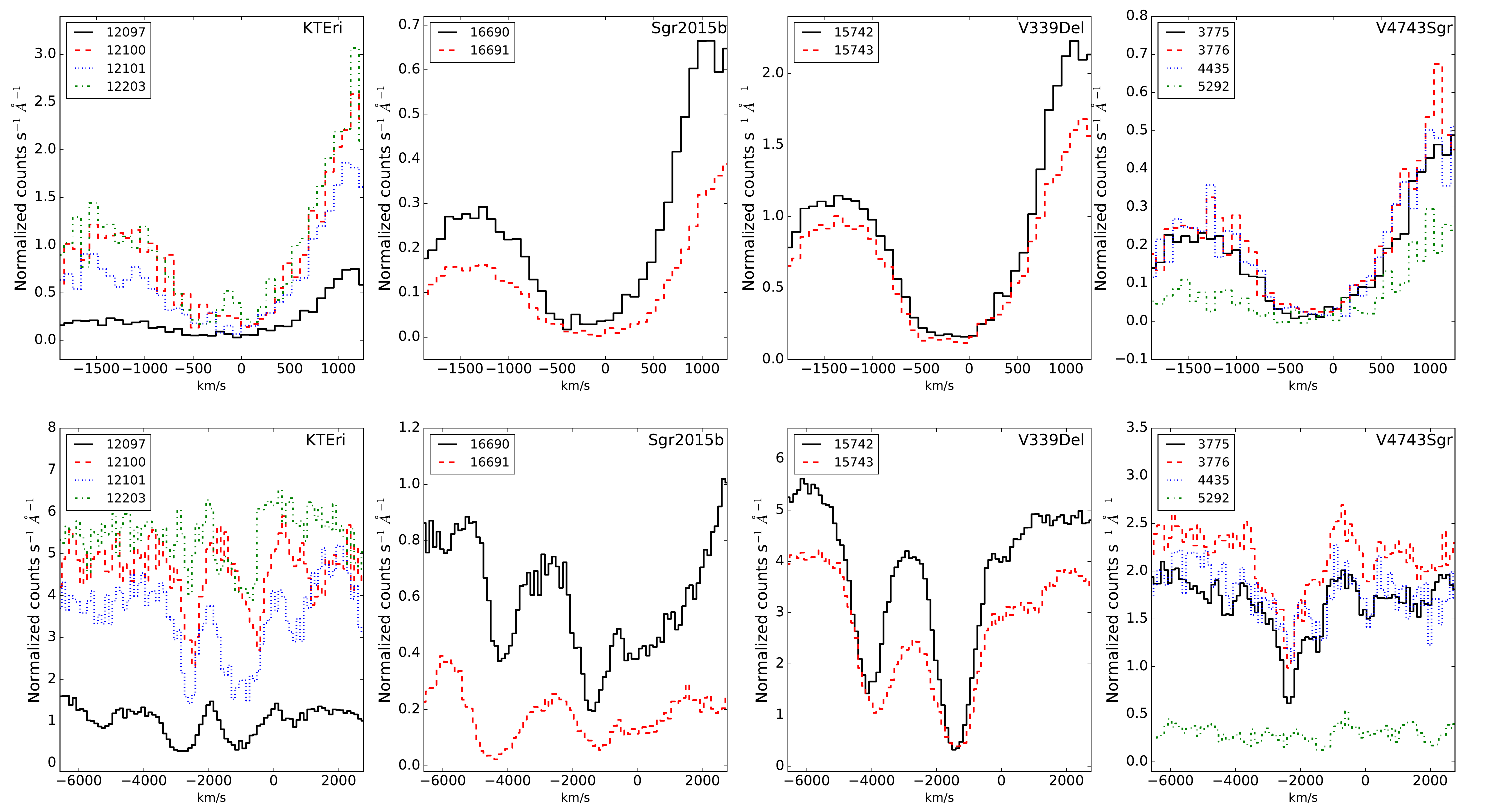} 
      \caption{ {\rm C}~{\sc ii} K$\alpha_{1}$ (top pannels) and {\rm C}~{\sc vi} K$\alpha$ (bottom pannels) absorption lines parametrized in velocity space for each source analyzed. }\label{fig_vel}
   \end{figure*}
 
\section{Results and discussions}\label{sec_dis}
 
        \begin{figure}
\includegraphics[scale=0.38]{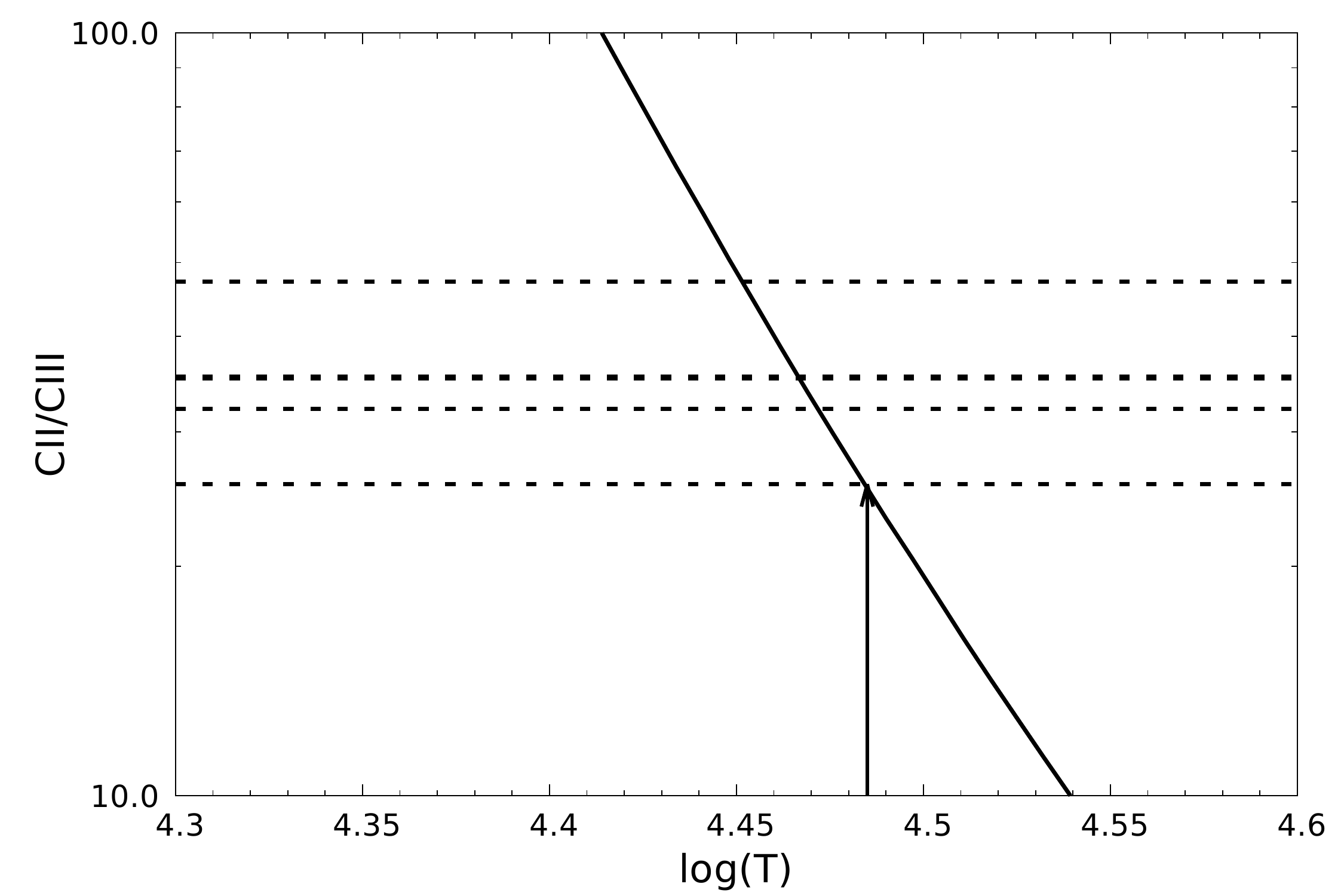} 
      \caption{ {\rm C}~{\sc ii}/{\rm C}~{\sc iii} ratio obtained from {\sc xstar} calculations assuming collisional ionization equilibrium plasma. Horizontal dashed lines correspond to the ratios listed in Table~\ref{tab3}. The arrow indicates the maximum temperature obtained. }\label{fig6}
   \end{figure}

We included the corrected cross-sections from \citet{has10} in the {\tt ISMabs}\footnote{\url{https://heasarc.gsfc.nasa.gov/xanadu/xspec/models/ismabs.html}} model in order to estimate {\rm C}~{\sc ii} and {\rm C}~{\sc iii} column densities in the SSS spectra listed in Table~\ref{tab1}. Figure~\ref{fig4} shows the best-fit results using the {\tt ISMabs} model for the {\rm C}~{\sc ii} K-$\alpha$ wavelength region (42.5--43.3 \AA). In each panel, black data points correspond to the observation. The model before the atomic data corrections and after the correction are indicated (red dashed and blue solid lines, respectively). In all cases we obtained a data fitting improvement of $\Delta \chi^{2} > 20$. Table~\ref{tab3} lists the column densities obtained for each observation.  Due to the elemental abundance enrichments in the ejected material from the respective novae explosion through the mixing process \citep{kel13}, differences between column densities for observations performed at different epochs are expected if the absorbing material is intrinsic to the source. Figure~\ref{fig5} shows a comparison between the column densities obtained from the {\tt ISMabs} model. It is clear from the plot that the column densities tend to agree for different observations of the same source. Differences between sources, on the other hand, are expected due to the density distribution of the ISM gas along the Galaxy \citep{rob03,kal09,nic16a,nic16b,gat18a}.

Another possibility is an origin in the SSS atmosphere which usually shows multiple absorption features \citep{ori12,nes12,nes13}. However, such atmospheres require high temperatures (> 0.6MK), for which we will not find {\rm C}~{\sc i}, {\rm C}~{\sc ii} and {\rm C}~{\sc iii} ions   \citep{nes09,rau10,van12,rau16}.  It is important to compare the 1$\sim$2 months observation separation time in our sample with the novae evolution time-scale, which can vary from months to years \citep{sch11}. For example, \citet{nes07} modeled the X-ray high-resolution spectra of the RS Ophiuchi novae, a source that shows a notable evolution in both, the continuum and the emission/absorption features, within months \citep[see Figure 1 in][]{nes09}. The model used by \citet{nes07} included a component for the ISM local absorption and a second component to model the circumstellar material intrinsic to the source. The best-fit requires that the oxygen contribution from the circumstellar material disappears around day 54 after the outburst, probably due to photoionization of the local gas by the radiation field. 

Figure~\ref{fig_vel} shows a comparison between {\rm C}~{\sc ii} K$\alpha_{1}$ (top pannels) and {\rm C}~{\sc vi} K$\alpha$ (bottom pannels) absorption lines parametrized in velocity space for each source analyzed in this work. The {\rm C}~{\sc vi} K$\alpha$ absorption line has been identified as intrinsic to the source previously \citep{nes03,pet05,nes07,nes12,van12}. It is clear from the plot that the {\rm C}~{\sc ii} K$\alpha_{1}$ remains at the same wavelength while there are no {\rm C}~{\sc vi} K$\alpha$ absorption features at rest wavelength. Also, some high-resolution X-ray novae spectra have shown P Cygni profiles \citep{nes07,ori13} which we have not identified in the analyzed spectra. 

It is important to note that there is no a reliable method to determine the total amount of carbon emitted in X-ray ejecta, in order to compare with the ISM abundance.  Theoretical estimation, such as \citet{rau10}, depends on multiple factors including the composition of the accreted material, nuclear burning products, composition of the white dwarf and the amount of mixing of white dwarf material into the ejecta.

Table~\ref{tab3} also list the {\rm C}~{\sc ii}/{\rm C}~{\sc iii} ratios for those sources for which both column densities can be estimated. It is clear that {\rm C}~{\sc ii} dominates in all cases. In this sense, previous analysis using {\it Herschel} Galactic observations show that {\rm C}~{\sc ii} constitutes the main carbon reservoir along the lines of sight in most cases \citep{pin14,ger15}. The ion fractions depend on the physical state of the plasma. We used the {\sc xstar}\footnote{\url{https://heasarc.gsfc.nasa.gov/lheasoft/xstar/xstar.html}} code to estimate the maximum temperature of the gas assuming collisional ionization equilibrium \citep[see][]{gat18a}.  Figure~\ref{fig6} shows the {\rm C}~{\sc ii}/{\rm C}~{\sc iii} ratio obtained from the {\sc xstar} calculations. Horizontal dashed lines correspond to the ratios listed in Table~\ref{tab3} while the vertical arrow indicates the maximum temperature derived. We found $T_{max}< 3.05\times10^{4}$ K.

\section{Conclusions and summary}\label{sec_con}

We have performed an analysis of the C K-edge using high-resolution {\it Chandra} spectra of four SSS. The instrumental features due to a C layer in the camera prevent the analysis of {\rm C}~{\sc i} absorption. We have detected all three resonances of the {\rm C}~{\sc ii} K$\alpha$ in 11 observations as well as the {\rm C}~{\sc iii} K$\alpha$ and K$\beta$ in 7 and 5 observations, respectively. We used the astronomical observations in order to perform a benchmarking of the atomic data computed by \citet{has10}. We have included these corrected cross-sections in the {\tt ISMabs} X-ray absorption model. Using the improved atomic data we estimated the {\rm C}~{\sc ii} and {\rm C}~{\sc iii} column densities for each observation. While high-ionized lines such as {\rm C}~{\sc vi} K$\alpha$ show significant shifts between different observations and different sources, the {\rm C}~{\sc ii} and {\rm C}~{\sc iii} wavelength positions are consistent. The absence of physical shifts for the absorption lines, the lack of variability for the column densities between different observations and the low temperatures associated to these ions compared to X-ray novae typical atmosphere temperature support our conclusion about an ISM origin of the absorption lines identified in the spectra. From the ratios of {\rm C}~{\sc ii}/{\rm C}~{\sc iii} column densities, we found $T_{max}< 3.05\times10^{4}$ K, which corresponds to the so-called warm component of the ISM.
   
\section*{acknowledgements}\label{sec_con}
 We are grateful to the referee, Dr. Lia Corrales, for the careful reading of our manuscript and the valuable comments that led to improvements of its scientific content.

\bibliographystyle{mnras}

\end{document}